\documentclass[twoside]{article}
\usepackage{fleqn,espcrc2,epsfig}

\def\be{\begin{equation}}
\def\ee{\end{equation}}
\def\bea{\begin{eqnarray}}
\def\eea{\end{eqnarray}}

\def\nn{\nonumber}

\newcommand{\AmS}{{\protect\the\textfont2
  A\kern-.1667em\lower.5ex\hbox{M}\kern-.125emS}}

\title{Mixmaster Ho\v{r}ava-Witten Cosmology}

\author{Mariusz P.~D\c{a}browski\address{Institute of Physics,
University of Szczecin, Wielkopolska 15, 70-451 Szczecin,
Poland\\E-mail: mpdabfz@uoo.univ.szczecin.pl}}

\begin{document}

\begin{abstract}
We discuss various superstring effective actions and,
in particular, their common sector which leads to the so-called
pre-big-bang cosmology (cosmology in a weak coupling limit of heterotic
superstring). Then, we review the main ideas of the Ho\v{r}ava-Witten
theory which is a strong coupling limit of heterotic superstring theory.
Using the conformal relationship between these two theories we
present Kasner asymptotic solutions of Bianchi type IX geometries
within these theories and make predictions about possible
emergence of chaos. Finally, we present a possible method of
generating Ho\v{r}ava-Witten cosmological solutions out of the well-known general
relativistic or pre-big-bang solutions.
\end{abstract}

\maketitle

\section{Introduction}

There are various superstring effective
theories (type I, type IIA, type IIB and heterotic $E_8 \times E_8$ or $SO(32)$)
which can be represented by effective actions with suitable field
equations which generalize general relativity. In view of duality symmetry the
superstring effective actions are not necessarily the right description of physics
at strong coupling, where string coupling parameter $g_s = \exp{(\phi/2)} \rightarrow \infty$.
It appears that at strong coupling regime the physics is 11-dimensional and can be described
by M-theory with its low-energy limit -- 11-dimensional supergravity.
One of the proposals for M-theory is Ho\v{r}ava-Witten theory.

The main issue of this talk is to discuss the question about the
implications of superstring/M-theory onto the evolution of the
universe.

\section{Cosmology of the common sector (pre-big-bang)}

Let us start with the presentation of superstring effective
actions. The first action under consideration is type {\bf IIA superstring} effective action.
It has $N=2$ supersymmetries of opposite chirality and reads
\cite{superjim}
\begin{eqnarray}
\label{IIA}
S_{\rm IIA} = \frac{1}{2\lambda_s^8} \left\{
\int d^{10}x \sqrt{-g_{10}}
\left[ e^{-\phi}  \left( R_{10} +
\left( \nabla \phi \right)^2  \right. \right. \right. \nonumber
\\ \left. \left. \left.  - \frac{1}{12} H_3^2 \right)
-\frac{1}{4} F^2_2 -\frac{1}{48} ({F_4}')^2 \right] +\frac{1}{2} \int B_2 \wedge F_4 \wedge F_4 \right\}
\end{eqnarray}
where $R_{10}$ is the Ricci scalar curvature of the spacetime with metric
$g_{\mu\nu}$ and $g_{10} \equiv {\rm det}g_{\mu\nu}$, $\lambda_s^2 = \alpha'$ is the fundamental
string length. Strings sweep out
geodesic surfaces with respect to the metric $g_{\mu\nu}$. The antisymmetric
tensor field strengths are defined by
$H_3=dB_2$, $F_2 =dA_1$, $F_4 =dA_3$ and $F'_4 =F_4 +A_1 \wedge H_3$,
where $A_{p}, B_{p}$ denote antisymmetric $p$--form potentials
and $d$ is the exterior derivative. The last term in
(\ref{IIA}) is a Chern--Simons term and is a necessary
consequence of supersymmetry \cite{polch}. The NS--NS sector
of the action contains the graviton, $g_{\mu\nu}$, the antisymmetric 2--form
potential, $B_2$, and the dilaton field, $\phi$. The RR sector contains antisymmetric
$p$--form potentials, where $p$ is odd. The NS--NS sector couples
directly to the dilaton, but the RR fields do not.

The bosonic type {\bf IIB superstring} effective theory has $N=2$
supersymmetries of the same chirality and its action reads
\begin{eqnarray}
\label{IIB}
S_{\rm IIB}=\frac{1}{2\lambda_s^8} \left\{
\int d^{10}x \sqrt{-g_{10}} \left[ e^{-\phi} \left( R_{10}+\left(
\nabla \phi \right)^2 \right. \right. \right. \nonumber \\ \left. \left. \left.
- \frac{1}{12} (H_3^{(1)})^2 \right)
 -\frac{1}{2} \left( \nabla \chi \right)^2 -
\frac{1}{12}(H_3^{(2)}+ \chi H_3^{(1)})^2  \right. \right. \nonumber \\
\left. \left. - \frac{1}{240} ({F_5})^2  \right]
+\int A_4 \wedge H_3^{(2)} \wedge H_3^{(1)} \right\} .
\end{eqnarray}
Here the bosonic massless excitations arising in the NS--NS sector
are the dilaton, $\phi$, the metric,  $g_{\mu\nu}$,
and the antisymmetric, 2--form potential, denoted here by $B^{(1)}_{\mu\nu}$.
The RR sector contains a scalar axion field,
$\chi$, a 2--form potential,  $B^{(2)}_{\mu\nu}$
(we dropped a 4--form potential, $D_{\mu\nu\rho\sigma}$ here) and
the RR field strengths are defined by
$H_3^{(2)} =dB_2^{(2)}$ and $F_5 =dA_4 +B_2^{(2)} \wedge H_3^{(1)}$.

In heterotic superstring theories supersymmetry is imposed only in
the right-moving sector so these theories are $N=1$
supersymmetric. Quantization of the left-moving sector requires
the gauge groups to be either $SO(32)$ or $E_8 \times E_8$ and the
choice of the group depends on the boundary conditions.
The {\bf heterotic superstring} effective action reads
\bea
\label{het}
S_{\rm H}=\frac{1}{2\lambda_s^8} \int d^{10} x
\sqrt{-g_{10}} e^{-\phi}
\left[ R_{10}+\left( \nabla \phi \right)^2 \right. \nonumber \\
\left. -\frac{1}{12} H_3^2 -\frac{1}{4} F_2^2 \right]
\eea
where $F_2^2$ is the field strength corresponding to
the gauge groups ${\rm SO}(32)$ or ${\rm E}_8\times {\rm E}_8$ and $H_3 = dB_2$ is the
field strength of a 2-form potential, $B_2$.

The last of the superstring theories is an open string theory or type I theory.
Because of the obvious reason (left and right moving sectors must be the same) it has $N=1$
supersymmetry and the effective action reads
\bea
\label{I}
S_{\rm I} =\frac{1}{2\lambda_s^8} \int d^{10}x
\sqrt{-g_{10}}\left[ e^{-\phi}
\left( R_{10} +\left( \nabla \phi \right)^2 \right) \right.
\nonumber \\ \left. -\frac{1}{12}
H^2_3 -\frac{1}{4} e^{-\phi /2} F^2_2 \right]
\eea
where $F^2_2$ is the Yang--Mills field strength taking values in the
gauge group $G= {\rm SO(32)}$ and $H_3 =dB_2$ is the field strength of a
2--form potential, $B_2$. We note that this field strength
is not coupled to the dilaton field in this frame and since both actions (\ref{het}) and
(\ref{I}) have the same particle content this is the only difference between the two
theories.

It is not difficult to notice that all the above theories (\ref{IIA})-(\ref{I})
in the string frame have the common sector
which (except different coupling of $H_3$ in type I) is
\be
\label{com10}
S = \frac{1}{2\lambda_s^8} \int d^{10}x\sqrt{-g_{10}}~e^{-\phi
}\left[R_{10} + \left( \nabla \phi \right)^2 - \frac
1{12}H_3^2\right].
\ee
In particular, (\ref{com10}) represents a weak coupling limit ($g_s \rightarrow 0$) of
heterotic $E_8 \times E_8$ theory. It is also the zeroth--order expansion in both
the string coupling $g_s$ and the inverse string tension $\alpha'$.

The common sector (\ref{com10}), after a suitable dimensional
reduction to 4 dimensions on a constant Calabi-Yau manifold, gives the following
elementary cosmological solutions for flat isotropic Friedmann geometry (with no axion
$H_3 = 0$)
\bea
\label{pbbsolns}
a(t) & = & \mid t \mid^{\pm \frac{1}{\sqrt{3}}} ,\\
e^{\phi(t)} & = & \mid t \mid^{\pm \sqrt{3} - 1} \nonumber  ,
\eea
where $a$ is the scale factor and $\phi$ the dilaton.
These solutions led to the cosmological scenario which is called
pre-big-bang cosmology \cite{ven91,gave93a,gave93b}.
It is easy to notice that the solutions (\ref{pbbsolns}) admit a phase of
expansion for \textit{\ negative} times as well as for positive times (with the
singularity formally located at $t=0$). It is both a curvature singularity
(Ricci scalar diverges) and a string coupling singularity $g_s \rightarrow
\infty$.

From (\ref{pbbsolns}) we realize that there are four possible types of the evolution
for the scale factor together with four corresponding types of evolution
for the dilaton.
These with `-' sign in (\ref{pbbsolns}) will be numbered as {\bf 1} and {\bf 2}
while those with `+' sign in (\ref{pbbsolns}) will be numbered as {\bf 3}
and {\bf 4}. All of them are commonly called branches. Branches {\bf 1}
and {\bf 3} apply for negative times ($t < 0$) while branches {\bf 2} and {\bf 4}
apply for positive times ($t >0$). The four types of evolution are
connected by the underlying symmetry of string theory namely $T-duality$
(also called O(d,d) symmetry \cite{meissner}) which in the context of
isotropic cosmology is called scale factor duality (SFD). Its
mathematical realization is given by the relation which interchange
the scale factor and the dilaton, leaving field equations unchanged, i.e.,
\bea
\label{SFD}
a(t)  \Longleftrightarrow \frac{1}{a(t)}   ,\\
\phi(t)  \Longleftrightarrow  \phi(t) - \ln{a^6(t)}   .
\eea
SFD relates {\bf 1} and {\bf 3} or {\bf 2} and {\bf 4} whose domains are
either for $t < 0$ or $t > 0$. However, there is also time-reflection
symmetry
\begin{center}
\be
\label{timeref}
\hspace{2.5cm} t \Longleftrightarrow - t  ,
\ee
\end{center}
which together with SFD gives relation beetween {\bf 1} and {\bf 4} as
follows
\be
\label{tref}
a_1(t) = (-t)^{-\frac{1}{\sqrt{3}}} \Longleftrightarrow t^{\frac{1}{\sqrt{3}}} =
\frac{1}{a_4(-t)}   .
\ee
It is easy to show that for branch {\bf 1}
\be
\frac{\ddot{a}_1}{a_1} > 0  ,
\ee
which means that it describes inflation which undergoes without a {\it violation of
the energy conditions} and it is called superinflation. This comes form the fact that
there is only {\it kinetic term} for the dilaton in the action (\ref{com10}) and there
is no potential energy at all. Let us remind that standard inflation is
potential-energy-driven inflation. It is easy to notice that the branch {\bf 4} is deflationary, i.e.,
\be
\frac{\ddot{a}_4}{a_4} < 0  ,
\ee
and it describes standard {\it radiation-dominated} evolution. Branches
{\bf 1} and {\bf 4} are duality-related, though, they are divided by
the singularity of curvature and strong coupling. The solutions
with axion are qualitatively the same -- still there is a
curvature singularity and strong coupling singularity though a
bounce of the scale factor $a(t)$ appears \cite{ed94}.

There are some problems with pre-big-bang scenario. One of the
most interesting is that in the conformally related Einstein frame
the action (\ref{com10}) is the same as the Einstein relativity
minimally coupled to a scalar field (or stiff fluid pressure =
energy density) and the solutions for negative times $t < 0$ are
just collapsing solutions and no superinflation is present.
Another is the so-called "graceful-exit" problem which is a
possible physical mechanism for a transition from superinflation
to a radiation-dominated universe. An interesting issue appears if
one considers less symmetric geometries like homogeneous Bianchi
or Kantowski-Sachs type models. In that case a choice of the
antisymmetric tensor potential is unclear since one is possible to
make it timely ($B_{\mu\nu} = B_{\mu\nu}(t)$) or spatially ($B_{\mu\nu} = B_{\mu\nu}(x)$)
dependent. The first case (though makes energy momentum tensor
time-dependent only) cannot be admitted to some geometries and even if
it can, it may prevent isotropization of the universe for late
times which is observationally unfavourable.

\section{M-theory and Ho\v{r}ava-Witten theory}

M-theory is defined as strong coupling limit of superstring
theories. It is 11-dimensional and it has got as a weak coupling limit
11-dimensional supergravity with the action
\begin{eqnarray}
\label{supergr}
S_{SUGRA}= \frac{1}{16\pi G_{11}}
\left( \int d^{11}x \sqrt{-g_{11}} \left[ R_{11}
-\frac{1}{48}F_4^2 \right] \right. \nonumber \\
\left. + \frac{1}{6} \int A_3 \wedge F_4 \wedge F_4,
\right)
\end{eqnarray}
where $G_{11}$ is 11-dimensional Newton constant and the Chern--Simons term arises as a direct
consequence of the ($N=1$) supersymmetry.

It has been shown that the compactification of $N=1$,
$D=11$ supergravity on a circle, $S^1$, results in the type IIA supergravity theory
which was interpreted as the strongly coupled limit of the type IIA superstring
(with $N=2$ supersymmetries) in terms of an 11-dimensional theory. This
correspondence gave Ho\v{r}ava and Witten \cite{hw} the idea that one can also
compactify eleven-dimensional supergravity on a $S^1/Z_2$ orbifold (which is a unit
interval $I$) in order
to get a heterotic theory with only $N=1$ supersymmetry. Other words, they proved that the
10-dimensional $E_8 \times E_8$ theory results from an 11-dimensional theory compactified
on the orbifold $R^{10} \times S^1/Z_2$ in the same way as the type IIA theory results from
an 11-dimensional theory compactified on $R^{10} \times S^1$.
This identifies strongly coupled limit of heterotic $E_8 \times E_8$
theory as 11-dimensional supergravity compactified on an orbifold.
The action for such a theory reads as
\be
\label{SHW}
S = S_{SUGRA} + S_{YM},
\ee
where
\bea
\label{SYM}
S_{YM} = - \frac{1}{8\pi\kappa^2_{11}}
\left(\frac{\kappa_{11}}{4\pi}\right)^{\frac{2}{3}} \left\{ \int_{M^{(1)}_{10}}
\sqrt{-g_{10}} \left[ tr\left(F^{(1)}\right)^2 \right. \right. \nonumber \\
\left. \left. - \frac{1}{2}
trR^2 \right] + \int_{M^{(2)}_{10}}
\sqrt{-g_{10}} \left[ tr\left(F^{(2)}\right)^2 - \frac{1}{2}
trR^2 \right] \right\} .
\eea
The action (\ref{SYM}) is composed of the two $E_8$ Yang-Mills theories on 10-dimensional
orbifold fixed planes -- manifolds $M^{(i)}_{10}$ (i = 1,2) and $F^{(i)}$ are the two
gauge field strengths.

\section{Ho\v{r}ava-Witten cosmological solutions}

For further cosmological investigations we can compactify Ho\v{r}ava-Witten
models on a Calabi-Yau deformed
manifold $X$ with orbifold coordinate to make decomposition $M^{11} = M^4 \times X
\times S^1/Z_2$. It is important that the size of the orbifold is much bigger than the
radius of the Calabi-Yau space and we can discuss 5-dimensional effective
theory with the action \cite{lukas}
\bea
\label{SHWg}
S = \int_{M_5} \sqrt{-g_5}
\left( \frac{1}{2} R - \frac{1}{2}\left( \nabla \phi \right)^2 - \frac{1}{6} \alpha_0^2
e^{-2\sqrt{2} \phi} \right) \nonumber \\
\mp \sqrt{2}
 \sum_{i=1}^{2} \int_{M^{(i)}_4} \sqrt{-\tilde{g}_4} \alpha_0 e^{-\sqrt{2}\phi},
\eea
where $M^{(1)}_4, M^{(1)}_4$ are orbifold fixed planes, $\phi = 1/\sqrt{2} \ln{V}$
is a scalar field (dilaton) which parametrizes the radius of Calabi-Yau
space and $\tilde{g}_{ij}, i,j = 0,1,2,3$ is the pull-back of
5-dimensional metric onto $M^{(1)}_4$ and $ M^{(1)}_4$. In the
action (\ref{SHWg}) we dropped other important fields like p-form fields,
gravitini, RR scalar and fermions.

The effective field equations for the action (\ref{SHWg}) are
($i,j = 0,1,2,3, \mu,\nu = 0,1,2,3,5$)
\begin{eqnarray}
 R_\mu ^\nu   =  \nabla _\mu \phi \nabla ^\nu \phi + \frac{\alpha_0^2}{9}
g_{\mu}^{\nu} e^{-2\sqrt{2}\phi}
+ \sqrt{2}\alpha_0
e^{-\sqrt{2}\phi}
\nonumber \\
\times \sqrt{\frac{\tilde{g}}{g}} \tilde{g}^{ij} \left[
g_{i\mu} g_j^{\nu} - \frac{1}{3} g_{\mu}^{\nu} g_{i\sigma} g_{j}^{\sigma}
\right] \left[ \delta(y) - \delta(y - \pi \lambda) \right]
\end{eqnarray}
\begin{eqnarray}
\frac{1}{\sqrt{-g}} \partial_\mu \left( \sqrt{-g} \partial^{\mu} \phi \right)
 =  - \frac{\sqrt{2}}{3} \alpha_0^2 e^{-2\sqrt{2} \phi} \nonumber \\ + 2 \alpha_0
\sqrt{\frac{\tilde{g}}{g}} e^{-\sqrt{2}\phi} \left[ \delta(y) - \delta(y - \pi \lambda)
\right].
\label{eom}
\end{eqnarray}
In (\ref{eom}) $y \equiv x^5 \in [-\pi \lambda, \pi \lambda]$ is a
coordinate in the orbifold direction and the orbifold fixed planes are at $y = 0, \pi
\lambda$. $Z_2$ acts on $S^1$ by $y \to -y$. The terms involving delta functions
arise from the stress energy on the boundary planes.

Before going further we present the form of 11-dimensional metric
for our cosmological solutions which is
\be
ds_{11}^2 = e^{-\frac{2\sqrt{2}}{3}\phi}
g_{\mu\nu}dx^{\mu}dx^{\nu} + e^{\frac{\sqrt{2}}{3}\phi}
\Omega_{mn}dy^{m}dy^{n}  ,
\ee
and $m,n = 6, \ldots, 11$ so that the last term is simply
Calabi-Yau metric. The 5-dimensional metric is given by
\be
ds_5^2 = g_{\mu\nu}dx^{\mu}dx^{\nu} = -N^2(\tau,y)d\tau^2 + ds_3^2 + d^2(\tau,y)dy^2
,
\ee
and our main task in this paper is to consider the most general
form of the 3-metric of homogeneous type Bianchi IX (or Mixmaster)
\begin{equation}
ds_3^2=a^2(\tau,y)(\sigma ^1)^2+b^2(\tau,y)(\sigma ^2)^2+c^2(\tau,y)(\sigma ^3)^2,
\end{equation}
where the orthonormal forms
$\sigma ^1,\sigma ^2,\sigma ^3$ are given by
\begin{eqnarray}
\sigma ^1 &=&\cos {\psi }d\theta +\sin {\psi }\sin {\theta }d\varphi ,
\\
\sigma ^2 &=&\sin {\psi }d\theta -\cos {\psi }\sin {\theta }d\varphi ,
\\
\sigma ^3 &=&d\psi +\cos {\theta }d\varphi ,
\end{eqnarray}
and the angular coordinates $\psi ,\theta ,\varphi $ span the
following ranges,
\begin{equation}
0\leq \psi \leq 4\pi ,\hspace{0.5cm}0\leq \theta \leq \pi
,\hspace{0.5cm}%
0\leq \varphi \leq 2\pi .
\end{equation}
One should notice that under a choice $\sigma^1 = dx^1, \sigma^2 = dx^2, \sigma^3 = dx^3$
one gets Bianchi type I flat geometry.

Similarly as in \cite{lukas,reall} we will look for separable
solutions of the form
\bea
\label{separable}
N(\tau,y) &=& n(\tau) \tilde{a}(y) , \nonumber \\
a(\tau,y) &=& \alpha(\tau) \tilde{a}(y) , \nonumber \\
b(\tau,y) &=& \beta(\tau) \tilde{a}(y) , \nonumber \\
c(\tau,y) &=& \gamma(\tau) \tilde{a}(y) ,  \\
d(\tau,y) &=& \delta(\tau) \tilde{d}(y) , \nonumber \\
V(\tau,y) &=& e^{\sqrt{2}\phi(\tau,y)} = \varepsilon(\tau) \tilde{V}(y) , \nonumber .
\eea
In fact $\alpha(\tau), \beta(\tau), \gamma(\tau)$ are vorldvolume
scale factors and $\delta(\tau)$ is an orbifold scale factor. It
appears that the suitable equations of motion are fully separable
into the orbifold-dependent part and spacetime-dependent part
provided \cite{lukas,3}
\be
n(\tau) = 1, \hspace{1.cm} \delta(\tau) = \varepsilon(\tau),
\ee
where the first condition is simply the choice of the lapse
function, while the second tells us that Calabi-Yau space is
tracking the orbifold. One can show that orbifold-dependent part
can be solved by
\bea
\tilde{a} & = & a_0 H^{1/2}(y) ,\nn \\
\tilde{d} & = & d_0 H^2(y) ,\nn \\
\tilde{V} & = & d_0 H^3(y) ,
\eea
where
\bea
H(y) & = & \frac{\sqrt{2}}{3} \alpha_0 \mid y \mid + h_0  ,\\
H''(y) & = & \frac{2\sqrt{2}}{3} \alpha_0 \left[ \delta(y) - \delta(y - \pi \lambda)
\right],
\eea
and we have applied
\be
\mid y \mid ' = \epsilon(y) - \epsilon(y - \pi \lambda) - 1 ,
\ee
so that
\be
\mid y \mid '' = 2 \delta(y) - 2 \delta(y - \pi \lambda) ,
\ee
(factor 2 comes from the fact that $y$ is periodic) and
\bea
\epsilon(y) = 1 \hspace{0.5cm} if \hspace{0.5cm} y \ge 0 ,\\
\epsilon(y) = -1 \hspace{0.5cm} if \hspace{0.5cm} y < 0 .
\eea
After all these substitutions one can write down 5-dimensional
metric in the form
\bea
\label{ds5H}
ds_5^2  =  -a_0^2 H(y) d\tau^2 + a_0^2 H(y) \left[ \alpha^2(\tau)(\sigma ^1)^2
\right. \nonumber \\ \left. +
\beta^2(\tau)(\sigma ^2)^2 + \gamma^2(\tau,y)(\sigma ^3)^2 \right]
  +  d_0^2 H^4(y) \delta^2(\tau) dy^2 .
\eea
Elementary solutions of Ho\v{r}ava-Witten theory for Friedmann
flat geometry analogous to (\ref{pbbsolns}) in pre-big-bang
cosmology are given by taking one scale factor $\bar{a} = \alpha = \beta = \gamma$
in (\ref{ds5H}) and read
\bea
\label{hwsolns}
\bar{a}(\tau) & = & \mid \tau \mid^{p_{\mp}}, \hspace{0.5cm}p_{\mp} = 3/11 \mp
4/11\sqrt{3}, \nn \\
\delta(\tau) & = & \mid \tau \mid^{q_{\mp}}, \hspace{0.5cm}q_{\pm} = 2/11 \pm
4\sqrt{3}/11,
\eea
for the worldvolume and orbifold respectively.
From (\ref{hwsolns}) one can easily deduce that there are four
types of evolution of the worldvolume $M^4$ and the orbifold,
namely: both worldvolume and orbifold contracts, both worldvolume
and orbifold expand, worldvolume contracts while orbifold expands
(superinflationary) and worldvolume expands while orbifold contracts.
The former case corresponds to superinflation while the latter to
standard radiation-dominated evolution in pre-big-bang scenario.

\section{Kasner asymptotics of Mixmaster Ho\v{r}ava-Witten and
pre-big-bang solutions}

Using the relationship between {\it weakly} (pre-big-bang (\ref{com10})) and {\it strongly}
(Ho\v{r}ava-Witten (\ref{SHWg})) coupled heterotic string theories \cite{lidsey} we study
the problem of the emergence of chaotic oscillations in Mixmaster cosmologies based on
these theories. In particular, we discuss Kasner asymptotic
states (anisotropic solutions of zero curvature to the field equations) of homogeneous
Bianchi type IX geometries in these string cosmologies. In order to present the time-dependent
part of the Ho\v{r}ava-Witten field equations we use a new time coordinate \cite{1}
\begin{equation}
d\eta =\frac{d\tau}{\alpha\beta\gamma\delta}  .
\end{equation}
From now on we will use the notation $(...),_{\eta} = d/d\eta$. To
further simplify the equations we additionally define
\begin{equation}
\tilde{\alpha} = \ln{\alpha}\hspace{0.2cm}\tilde{\beta} = \ln{\beta}
\hspace{0.2cm}\tilde{\gamma} = \ln{\gamma }\hspace{0.2cm}
\tilde{\delta} = \ln{\delta} ,
\end{equation}
so that we get
\bea
\label{fe}
\left(\tilde{\alpha} + \tilde{\beta} +\tilde{\gamma} + \tilde{\delta} \right) _{, \eta \eta }
 +  \frac{1}{2} \tilde{\delta}_{,\eta}^2
= 2\left(\tilde{\alpha}_{,\eta}\tilde{\beta}_{,\eta} +
\tilde{\alpha}_{,\eta}\tilde{\gamma}_{,\eta} \right. \nonumber \\ \left. +
\tilde{\beta}_{,\eta}\tilde{\gamma}_{,\eta}\right)
 +  2\left(\tilde{\alpha}_{,\eta} + \tilde{\beta}_{,\eta } + \tilde{\gamma}_{,\eta}\right)
\tilde{\delta}_{,\eta },
\eea
\bea
2\tilde{\alpha}_{,\eta \eta } &=& \left[ \left(
\beta^2-\gamma^2 \right)^2-\alpha^4 \right]\delta^2, \\
2\tilde{\beta}_{,\eta \eta} &=&\left[ \left(
\alpha^2-\gamma^2\right)^2-\beta^4 \right]\delta^2, \\
2\tilde{\gamma}_{, \eta \eta } &=&\left[ \left(
\beta^2-\gamma^2\right)^2-\alpha^4 \right]\delta^2,  \\
\tilde{\delta}_{,\eta\eta} &=& 0 .
\eea
The important point is that pre-big-bang field equations are the
same except the constraint which now takes the form ($\tilde{\delta} = -
\phi$)
\bea
\label{kasnerfe}
\left(\tilde{\alpha} + \tilde{\beta} +\tilde{\gamma} \right) _{, \eta \eta }
- \tilde{\delta}_{,\eta}^2
= 2\left(\tilde{\alpha}_{,\eta}\tilde{\beta}_{,\eta} +
\tilde{\alpha}_{,\eta}\tilde{\gamma}_{,\eta} + \tilde{\beta}_{,\eta}\tilde{\gamma}_{,\eta}\right)
\nonumber \\
- 2\left(\tilde{\alpha}_{,\eta} + \tilde{\beta}_{,\eta } + \tilde{\gamma}_{,\eta}\right)
\phi_{,\eta } .
\eea
The Kasner asymptotic solutions are of the type
\begin{equation}
\label{kasner}
\alpha(\tau) = \alpha_0 \tau^{p_1},  \beta(\tau) = \beta_0 \tau^{p_2},
\gamma = \gamma_0 \tau^{p_3}, \delta = e^{\phi} = \delta_0
\tau^{p_4},
\end{equation}
where $\alpha, \beta, \gamma$ are scale factors, $\phi$ is the scalar
field, $\alpha_0, \beta_0, \gamma_0, \delta_0$
constants and $p_1, p_2, p_3, p_4$ the so-called Kasner indices (it is obvious that the
`fourth' Kasner index $p_4$ refers to the scalar field).

It is well-known that the approach to a singularity in Bianchi
type IX models of the universe happens through a sequence of the
Kasner-to-Kasner transitions (Mixmaster oscillations). These
transitions can be described as the replacements of the Kasner
indices of the type
\be
p_i' = p_i'(p_i)   .
\ee
The main problem is to establish whether the number of these
replacements will be finite (no chaos) or infinite (chaos). It is
easy to notice from the field equations (\ref{fe}) that
these transitions are possible only if the terms of the type
\bea
\label{increase}
\alpha^4\delta^2 \propto \tau^{(2p_1 + p_4)} = \tau^{(1 + p_1 - p_2 - p_3)}   \nonumber,\\
\beta^4\delta^2 \propto \tau^{(2p_2 + p_4)} = \tau^{(1 + p_2 - p_3 - p_1)}   ,\\
\gamma^4\delta^2 \propto \tau^{(2p_3 + p_4)} = \tau^{(1 + p_3 - p_1 - p_2)}   \nonumber,
\eea
increase in the limit $\eta \to \infty$ ($\tau \to 0$). The
increase is possible in the regions which are marked in the
Figures 1 and 2. Fig. 1 corresponds to pre-big-bang while Fig. 2
corresponds to Ho\v{r}ava-Witten. However, this is not enough to
answer the question about the infinite continuation of Kasner
transitions since the increase may, in general, be replaced by the
decrease of these indices in the marked regions either. What is
important is whether the increase (or the decrease) of the indices
refers to all of them and whether they all may have the same sign.
If they do have the same sign, the expansion becomes monotonic - a
singularity is reached and there is no chaos.

To establich that in both limits of the heterotic $E_8 \times E_8$ theory we obtain suitable
conditions for Kasner indices ("Kasner sphere") which are \cite{1,2}:
\begin{equation}
\label{weak}
p_1 + p_2 + p_3 + p_4 = 1, \hspace{0.5cm}
p_1^2 + p_2^2 + p_3^2 = 1
\end{equation}
for weak coupling limit (pre-big-bang) and \cite{3}
\begin{equation}
\label{strong}
p_1 + p_2 + p_3 + p_4 = 1, \hspace{0.5cm}
p_1^2 + p_2^2 + p_3^2 + \frac{3}{2}p_4^2 = 1
\end{equation}
for strong coupling limit. For further discussion it is important that both Kasner
asymptotics contain the isotropic solutions of Friedmann type as
special cases in their domains. For (\ref{weak}) these points are defined by
$p_1 = p_2 = p_3 = \pm 1/\sqrt{3}, p_4 = \mp \sqrt{3} + 1$ while for
(\ref{strong})
these points are defined by $p_1 = p_2 = p_3 = p_{\mp} = 3/11 \mp 4/11\sqrt{3},
p_4 = q_{\pm} = 2/11 \pm 4\sqrt{3}/11$. This means there is a nonzero
region in the parameter space where all indices can have the same
sign (e.g. all positive). Once it happens the chaotic scatterings
of the anisotropic Bianchi IX model stop which means that it is
impossible to approach singularity in a chaotic way.

\begin{figure}[t]
\centering
\leavevmode\epsfysize=5cm \epsfbox{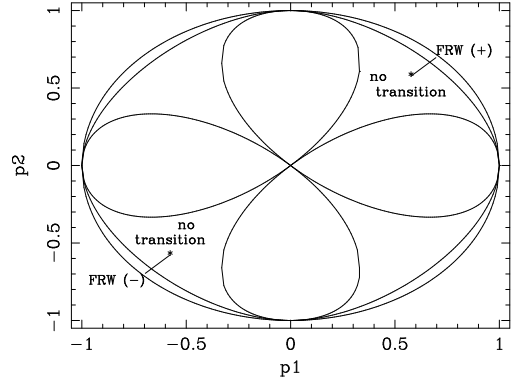}\\
\caption[]
{The range of Kasner indices $p_1$ and $p_2$ which fulfill the conditions
(\ref{increase}) for pre-big-bang cosmologies. The appearence of the
isotropic FRW (+) and (-) prevent chaotic oscillations in the neighbouring regions.}
\label{fig1}
\end{figure}

\begin{figure}[t]
\centering
\leavevmode\epsfysize=5cm \epsfbox{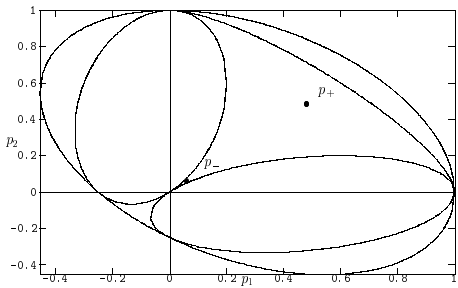}\\
\caption[]
{The range of Kasner indices $p_1$ and $p_2$ which fulfill the conditions
(\ref{increase}) for Ho\v{r}ava-Witten cosmologies.
The appearance of the isotropic Friedmann cases at $p_1 = p_2 = p_-$  and $p_1 = p_2 = p_+$
prevents chaotic oscillations in the shaded region that surrounds them.}
\label{fig2}
\end{figure}

\section{Weak coupling versus strong coupling -- how to generate
Ho\v{r}ava-Witten solutions}

In fact, there is a relationships between the pre-big-bang
(\ref{weak}) and Ho\v{r}ava-Witten (\ref{strong}) solutions and
the result of our previous section is an example of generation of solutions
from those known in one theory, into the other - in particular, into solutions of
Ho\v{r}ava-Witten theory \cite{lidsey}. This happens due to
conformal relation between the theories:
\be
g_{\mu\nu}^{E} = e^{-\phi} g_{\mu\nu}^{S} = e^{\frac{1}{2}\phi}
g_{\mu\nu}^{HW}  ,
\ee
where $E, S, HW$ refer to Einstein frame, string frame and
Ho\v{r}ava-Witten, respectively, and the relation is true,
provided 3-branes are given by the separable ansatz equations
(\ref{separable}). In particular, lots of exact solutions are
available for the Einstein frame (general relativity + stiff fluid
matter) - having them, one is able to generate Ho\v{r}ava-Witten
type solutions and study their properties.

Our conclusions are as follows. Ho\v{r}ava-Witten theory admits Mixmaster type
cosmology with Kasner type asymptotic solutions. Due to conformal relations one can
generate Ho\v{r}ava-Witten cosmological solutions (with separable 3-brane) of many types and
study their properties. Ho\v{r}ava-Witten Mixmaster cosmology (for truncated spectrum
of particles), similarly as pre-big-bang (truncated) cosmology
does not admit chaos.

Finally, one should mention some open issues in the topic. Firstly,
one should study non-separable ans\"atze for Ho\v{r}ava-Witten cosmology
and investigate how they relate to pre-big-bang.
Secondly, one
should study Mixmaster behaviour of cosmological models which
involve non-truncated spectrum of particles and inhomogeneities
\cite{damour1,damour2}.

\section*{Acknowledgments}

This work was supported by the Polish Research Committee (KBN) grant No 2 PO3B 105 16.

\end{document}